# Sustainable development-oriented campus bike-sharing site evaluation model: A case study of Henan Polytechnic University

Huimin Qi[a] ; Xianghong Li[a]*; Kai Yin[b]; Xiangnan Song[a]; Xufei Fang[a]

[a]*Department of Energy Science and Engineering, Henan Polytechnic University, Jiaozuo, China;*
[b]*School of traffic and transportation, Beijing Jiaotong University, Beijing, China*

**Abstract**: Promoting sustainable transportation options is increasingly crucial in the pursuit of environmentally friendly and efficient campus mobility systems. Among these options, bike-sharing programs have garnered substantial attention for their capacity to mitigate traffic congestion, decrease carbon emissions, and enhance overall campus sustainability. However, improper selection of bike-sharing sites has led to the growing problems of unsustainable practices in campus, including the disorderly parking and indiscriminate placement of bike-sharing. To this end, this paper proposes a novel sustainable development-oriented campus bike-sharing site evaluation model integrating the improved Delphi and fuzzy comprehensive evaluation approaches. Fourteen evaluation metrics are firstly selected from four dimensions: the user features, implementation and usage characteristics of parking spots, environmental sustainability, and social sustainability, through the combination of expert experience and the improved Delphi method. Then, the analytic hierarchy process and the entropy weight method are employed to determine the weights of the evaluation indices, ensuring a robust and objective assessment framework. The fuzzy comprehensive evaluation method is finally implemented to evaluate the quality of location selection. South Campus of Henan Polytechnic University is selected as a case study using the proposed evaluation system. This work contributes to the existing body of knowledge by presenting a comprehensive location selection evaluation system for campus bike-sharing, informed by the principles of sustainable development.

**Keywords**: sustainable development, bike-sharing, campus, location, Delphi method

*Corresponding Author
  *Email address:*lixianghong@hpu.edu.cn (Li Xianghong)

# 1. Introduction

## *1.1 Motivation*

In recent years, the concept of sustainable development has gained significant prominence across various sectors, driven by the increasing global awareness of environmental degradation, resource depletion, and the need for responsible growth. Among the array of sustainability initiatives, urban mobility and transportation systems have emerged as critical domains for fostering more sustainable living environments. One innovative solution that aligns with these objectives is the implementation of campus bike-sharing programs, which seek to promote eco-friendly transportation alternatives while addressing the challenges of urban congestion and pollution. Bike-sharing is an eco-friendly transportation operating model in the context of the sharing economy. It could not only fulfill the short-distance travel needs of residents and alleviates traffic congestion in urban areas, but also hold great significance in promoting the sustainable development of the urban environment(Hu et al., 2022). The campus, as a crucial element of urban functional zones(Sun et al., 2013), plays a significant role in expediting the area's urbanization while contributing to job creation, innovation, and the overall economic vitality of the city(Mohammed et al., 2022). Additionally, the campus actively promotes mixed land use in the region(Mohammed and Ukai, 2022). Campuses, as microcosms of larger urban environments, offer a unique setting to explore the potential of shared bicycle programs in fostering sustainable development. However, there are some problems of campus bike-sharing site selection in real-world application, ranging from poorly planned distribution of bike-sharing stations or an insufficient number of stations, to unreasonable location selection of legal parking sits. These have led to imbalances in the accessibility of bikes(He et al., 2022), random parking(Wu and Lei, 2019), and "difficulty picking and parking"(Gao et al., 2018), which impedes the sustainable development of urban system. Research showed that optimizing the spatial distribution of shared bicycles can promote the sustainable development of urban transportation systems(Caggiani et al., 2020; Chen et al., 2022; Croci and Rossi, 2014; Lu et al., 2018).

Considering the problems in the real-world application of campus bike-sharing site selection, it is vital to establish a comprehensive evaluation system for sustainable development of campus bike-sharing, and figure out the influencing factors and influencing mechanisms in the campus bike-sharing site selection. These are crucial for promoting the sustainable development of the campus, and making bike-sharing an appealing choice for short trips within the campus by strategically situating locations.

*1.2 Research Gap*

The current research on bike-sharing location selection primarily emphasizes the entire urban system(Chen et al., 2022; Macioszek and Cieśla, 2022), often overlooking the consideration of the campus environment. However, it is important to note that the campus transportation system differs significantly from the urban transportation system. Campus travel exhibits unique characteristics, such as a smaller scale, space constraints, trip distances, travel patterns, and pedestrian focus(Qin et al., 2023; Schnarre et al., 2022). When compared to the features of urban travel(Liu et al., 2022; Shen et al., 2022; Shen et al., 2022; Shi et al., 2021; Wang et al., 2021; Yin et al., 2018), campus travel is characterized by highly concentrated origins and destinations, as well as a distinct sense of time urgency. Travelers within campus environments opt for different modes of transportation depending on the level of time urgency they encounter(Jiageng et al., 2022). As a result, location development research conducted solely within the context of the entire urban environment cannot be directly applied to the specific circumstances and complexities of the campus setting. Therefore, this paper mainly focused on examining the location selection of campus bike-sharing taking the differences between campus and urban travel environments into account.

Apart from that, current research still lacks a thorough analysis of the factors that impact the selection of bike-sharing locations on campus from the perspective of sustainable development. The factors that can influence the sustainable development of campus bike-sharing location include: stakeholder engagement (e.g., students, faculty, staff, and campus administrators)(Mete et al., 2018), land use and space utilization(Eren and Katanalp, 2022), using features(Song et al., 2020), and environmental sustainability(Wang and Sun, 2022). By thoroughly analyzing these factors from the perspective of sustainable development, campus planners and administrators can make well-informed decisions regarding the location of bike-sharing. In turn, this will promote the campus transportation system to be more sustainable, user-friendly, and aligned with environmentally conscious practices(Wang et al., 2022; Yin et al., 2023).

Creating a comprehensive inventory of influencing factors is vital to improve the sustainable development of campus bike-sharing. However, the majority of existing studies concentrate solely on evaluating one or a few dimensions related to the site selection of campus bike-sharing, overlooking the broader picture. Cheng and Wei (2020) evaluates the bike-sharing spots selection from four aspects: accessibility to the destination, walkability to the transportation hub, flexibility of bicycle renting and returning, and the interest of the bike-sharing company. Additionally, other research has taken into account the user population and assessed the choice of bike-sharing locations based on three factors: urban life, user count, and transport network(C, 2017). The existing evaluation systems for bike-sharing typically assess the performance of campus bike-sharing system location selection based on single factors such as user feedback and satisfaction(Ricci, 2015), usage patterns and demands(Lin et al., 2020), and operational efficiency(Li et al., 2022). There are also studies that consider the implementation conditions of bike-sharing, Faghih-Imani et al. (2014) conducted a study where they utilized a multilevel statistical modeling approach to analyze the impact of bicycle infrastructure, land use, and built environment attributes on arrival and departure flows at the station level. There are also studies that consider the using features of bike-sharing, Faghih-Imani et al. (2014) visually depicted the bicycle arrival and departure rates for each station and discovered that the addition of a station has a more significant effect compared to increasing station capacity. García-Palomares et al. (2012) proposed a GIS-based method for determining station capacity and defining demand characteristics. Based on the principle of maximum service radius of parking spots and taking into consideration the nature of building services, Guo et al. (2020) developed an optimization model for the layout of campus bike-sharing parking spots. In this study, a thorough literature review of previous studies was conducted to determine the factors that impact the selection of bike-sharing locations (Table 1). However, none of these systems incorporate environmental factors into the evaluation framework. Moreover, it has been observed that the environment can significantly impact the extent of sustainable development in campus bike-sharing(Zhu et al., 2022). Therefore, this paper integrates the concept of sustainable development and incorporates environmental influences to construct a more comprehensive evaluation system for studying the quality of campus bike-sharing location selection.

Table 1 Influencing factors of bike-sharing location selection

| Indicator classification | Indicator | References for justification |
|---|---|---|
| User's characteristics | The proximity of home to stations | (Bachand-Marleau et al., 2012) |
| | User motivation | (Bachand-Marleau et al., 2012) |
| | Time urgency | (Jiageng et al., 2022) |
| | Search time | (Fricker and Gast, 2016) (Zhang et al., 2018) |
| Implementation conditions | Return time | (Zhang et al., 2018) |
| | Land use and built environment | (Faghih-Imani et al., 2014) |
| | Environmental benefits | (Zhang and Mi, 2018) |
| | Business benefits | (Choi and Choi, 2020) |
| Using features | Usage rates | (Fishman et al., 2013) |
| | Arrival and departure rates | (Faghih-Imani et al., 2014) |
| | Maximizing coverage | (García-Palomares et al., 2012) |
| | Station capacity | (García-Palomares et al., 2012) |
| | Maximum bicycle demand | (Duran-Rodas et al., 2021) |
| | Parking capacity | (Xu et al., 2019) |
| | Service distance | (Yan et al., 2020) |
| | Maximum service radius | (Guo et al., 2020) |
| Management level | Policy interventions | (Ye, 2022) |
| | Enterprise management measures | (Vogel et al., 2011) |
| | Campus management facilities | (Mete et al., 2018) |

There are certain shortcomings in evaluating the sustainable development of bike-sharing location on campus. A comprehensive evaluation system to assess the sustainability of bike-sharing on campus is presently lacking. To address these gaps, the objective of this study is to establish a comprehensive evaluation system that can identify the crucial factors influencing the sustainable development of bike-sharing location on campus.

*1.3 The Focus of This Study*

In this study, we aim to evaluate the quality of campus bike-sharing location selection from the perspective of sustainable development. Firstly, we select 30 initial evaluation metrics from perspectives of the user's characteristics, the implementation conditions of the parking points, the use characteristics of the parking points, environmental sustainability, and social sustainability. Secondly, delph method is further adopted to determine the list of key influencing factors, and the main indicators include time urgency, the environmental protection concept of users, and the compatibility with building plots. Thirdly, an evaluation model is developed based on the Analytic Hierarchy Process (AHP) and Entropy Weight Method.

Finally, South Campus of Henan Polytechnic University is selected as a case study using the proposed fuzzy comprehensive evaluation system.

Our work provides contributions from the following perspectives:

(1) From the perspective of sustainable development, the integration of sustainability indicators into the evaluation framework for campus bike-sharing could improve the overall sustainability of the campus transportation system.

(2) Introducing public participation into the Delphi method to construct the modified Delphi method that takes into account users' preferences in the evaluation system of campus bike-sharing location demonstrates the characteristics of public engagement.

(3) By adopting the perspective of campus decision-makers instead of that of corporations, a fuzzy comprehensive evaluation model was developed to assess the selection of campus bike-sharing sites. This model ensures the sustainability of the campus transportation system and promotes overall campus sustainability.

The remaining sections of this paper are organized as follows. Section 2 provides an overview of the relevant literature. Section 3 presents the evaluation indicators and methods for the selection of bike-sharing location on campus. Section 4 establishes the evaluation framework and provides detailed explanations of each indicator. Section 5 validates the rationality of the evaluation framework using the case of the South Campus of Henan Polytechnic University. Section 6 discusses the contributions, significance, limitations, and future work of this study.

## 2. Methodology

### 2.1 Model framework

This study employed a mixed-methods approach, combining qualitative and quantitative research methods. Fig. 1 illustrates the flowchart of the research process in this study. Firstly, a qualitative approach was utilized, including literature review and an improved Delphi method, to determine a set of candidate evaluation indicators. Subsequently, the Analytic Hierarchy Process (AHP) and Entropy Weight Method were employed to determine the subjective weights and objective weights of each indicator value, respectively. Finally, the Fuzzy Comprehensive Evaluation method was utilized to assess the quality of the bike-sharing location selection. In comparison to traditional methods of bike-sharing location selection optimization, the integration of the AHP, Entropy Weight Method, and Fuzzy Comprehensive Evaluation method in the evaluation optimization approach presents several benefits. This combined approach enables the decision-making process to incorporate multiple evaluation objectives within specific regions, effectively utilizing quantitative information. As a result, it mitigates potential errors stemming from inadequate considerations and enhances the accuracy of decision-making processes.

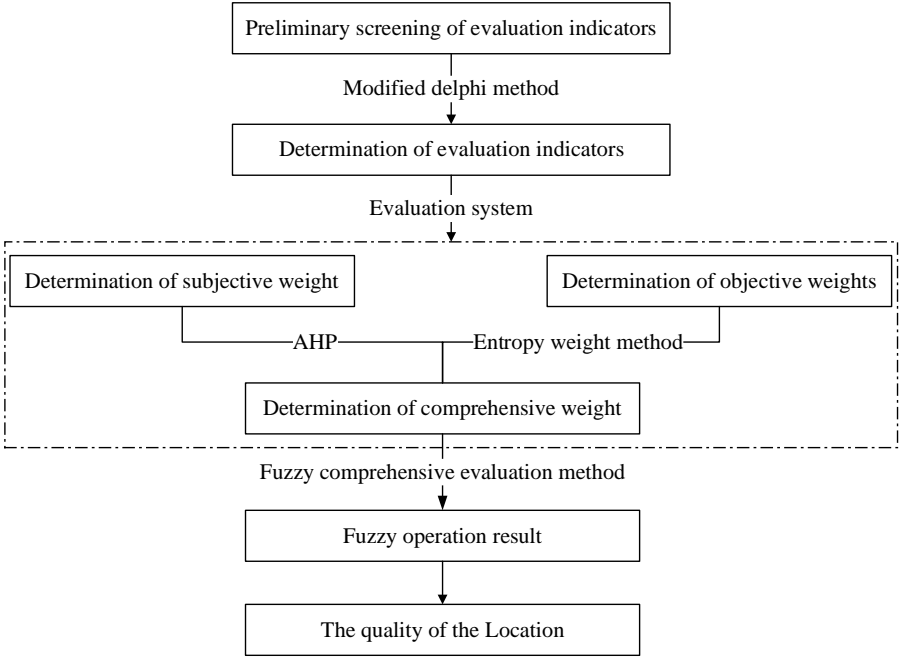

Fig. 1. The flow chart of the proposed methodology.

## 2.2 Evaluation indicators selection

### 2.2.1 Determination of evaluation indicators

Related studies have shown that the selection of bike-sharing parking spots on campus can be summarized as a multi-objective evaluation of the user's characteristics, the implementation conditions of the parking spots themselves, using features, management level, environmental sustainability, and social sustainability(Cheng and Wei, 2020; Yahya, 2017). The user's perspective refers to the factors that affect the user's use of bike-sharing, such as the average time of arrival of the user, the walking distance from the user to the parking point, time urgency, etc. The implementation conditions of the parking point itself mainly include factors such as the pedestrian flow around the parking point, the impact on business benefits, the impact on environmental benefits, and the impact on the surrounding traffic environment. The using features refer to some basic attributes of a parking spot, including floor area, usage rate, maximum parking capacity, etc. The management level is mainly considered from three perspectives: the state, enterprises, and campus, including the strength of national policies, improvement of enterprise management measures, and campus management facilities. The environmental sustainability is mainly considered from aspects such as landscape integration, compatibility with building plots, and optimization of spatial order. The social sustainability mainly includes factors such as the construction of corporate environmental protection culture, the promotion of corporate environmental protection concepts, the implementation of campus environmental education, and the environmental protection concepts of users.

Based on the above evaluation objectives(Kiptum et al., 2022; Pamucar et al., 2022; Xie et al., 2020; Xin et al., 2018), this study first conducted a comprehensive literature review of previous studies and preliminarily determined a list of alternative indicators for the location of bike-sharing stations on campus, as shown in Fig. 2. A total of 30 evaluation indicators were preliminarily identified as influencing factors for the location of bike-sharing stops.

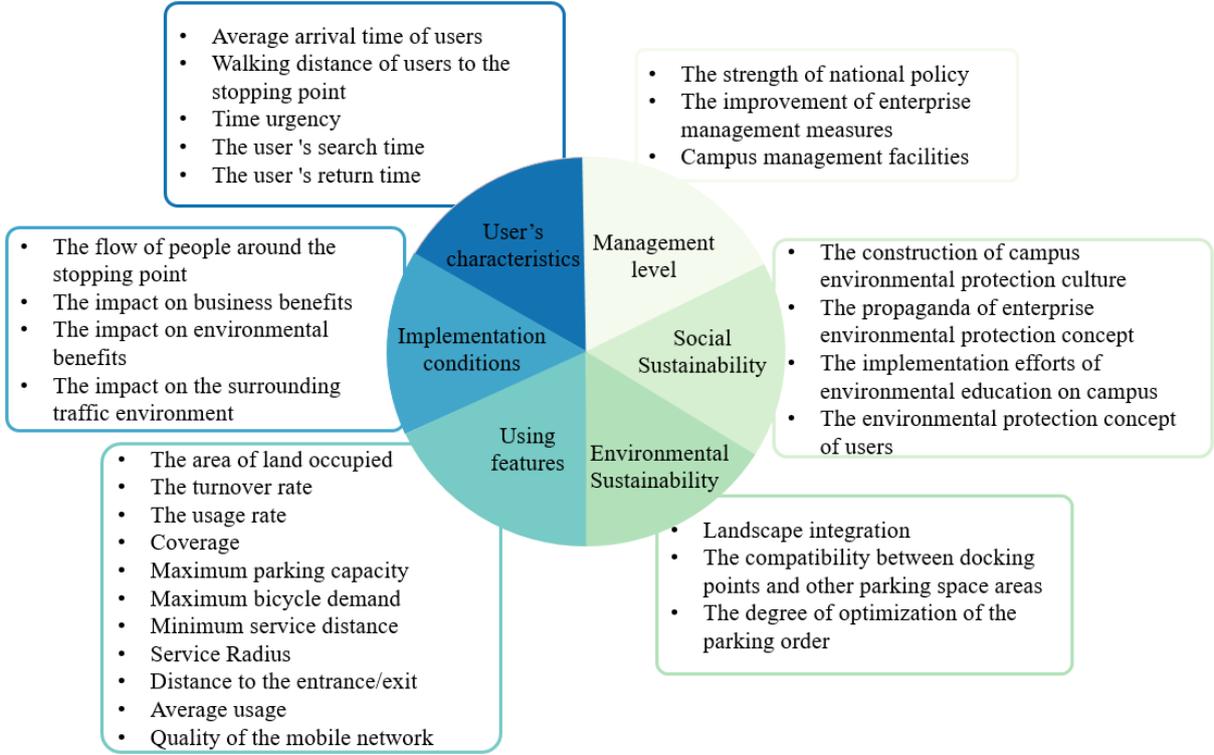

Fig.2 Alternative indicators for location selection of bike-sharing parking points on campus

*2.2.2 The improved Delphi method for screening indicators*

The university community—comprising students, faculty, and staff—serves as an engaged and accessible stakeholder group that directly experiences the campus bike-sharing program. This proximity fosters greater involvement and participation in the study, enabling researchers to gather comprehensive data on user preferences, usage patterns, and perceptions. The insights garnered from these stakeholders contribute to a more holistic understanding of the factors influencing the success and sustainability of the bike-sharing initiative.

(1) Identify expert advisors. 5 experts and 5 students from different work units and backgrounds (Table 2) were invited to review and refine the list of influencing factors. The invited experts mainly come from universities and research institutions, and each expert has in-depth research on shared travel; The invited students are college students who often use bike-sharing as a means of transportation.

Table 2 Respondents' information

| Number | Age | Gender | Work Affiliation | Position |
|---|---|---|---|---|
| 1 | 42 | female | Henan Polytechnic University | Professor |
| 2 | 47 | male | Nanjing University of Science and Technology | Professor |
| 3 | 30 | female | Technical University of Denmark | Doctor |
| 4 | 45 | male | Henan Polytechnic University | Professor |
| 5 | 43 | female | Transportation planning Design and Research Institute of Henan Province | Senior Engineer |
| 6 | 19 | female | Henan Polytechnic University | Student |
| 7 | 20 | male | Henan Polytechnic University | Student |
| 8 | 21 | female | Henan Polytechnic University | Student |
| 9 | 22 | male | Henan Polytechnic University | Student |
| 10 | 23 | male | Henan Polytechnic University | Student |

(2) A survey questionnaire was sent to all respondents via email and they were asked to express their opinions on the list of influencing factors. The received summary of opinions is sent back to all respondents, and they are then asked to modify their choices. This process was repeated two times until the final comprehensive evaluation system for the location of shared bicycle parking points on campus was approved by all respondents.

(3) Conduct statistical analysis on the data from the second round of the consultation questionnaire, which involves calculating metrics such as the mean, standard deviation, percentage of the maximum score, and coefficient of variation. After reviewing relevant literature(Quyên, 2014), this study has chosen three indicators, namely the percentage of maximum score, mean, and coefficient of variation, as the criteria for screening. The screening criteria are as follows: all evaluation indicators must satisfy the conditions of mean > 3.5, percentage of maximum score > 0.5, and coefficient of variation < 0.25. Among them:

The percentage of maximum score represents the ratio of experts who assign a score of 4 or 5 to a criterion out of the total number of experts. A higher percentage of maximum score signifies that a larger number of experts have awarded the criterion with the highest score, suggesting a greater influence and significance of the criterion on the selection of the location for sharing-bike stations.

$$K_i^{'} = m_i^{'} / m_i \tag{1}$$

In equation (1), $K_i'$ represents the percentage of maximum score, $m_i'$ represents the number of experts who give the maximum score to the i-th criterion, and $m_i$ represents the total number of experts participating in the evaluation of the i-th criterion.

Since this study involves expert consultants who are also end users, a modification was made to the Delphi method when calculating the percentage of maximum score. The scoring weight of planning and management experts was set at 0.8, while the scoring weight of end users was set at 0.2. This adjustment was made to ensure a more precise screening process in the evaluation system, considering that end users evaluate the docking points solely from their own perspective.

The mean is utilized to indicate the average score assigned by experts to each criterion, and a larger mean similarly signifies a higher degree of importance for that criterion.

$$\overline{E_i} = \sum_{j=1}^{5} E_j m_{ij} / d, i = 1, 2, \cdots n \qquad (2)$$

In equation (2), $E_j$ represents the magnitude (1, 2, 3, 4, 5) indicating the level of evaluation for the impact of each criterion on the selection of shared bicycle locations. $m_{ij}$ represents the number of experts who assigned the j-th level to the i-th evaluation criterion. "n" represents the number of criteria, and "d" represents the number of experts.

The coefficient of variation is employed to indicate the level of variability in the experts' evaluations of a criterion. A smaller coefficient of variation signifies a higher degree of agreement among the experts' recommendations.

$$\sigma_i = [\sum_{j=1}^{5} m_{ij}(E_j - \overline{E_i})^2 / (d-1)]^{\frac{1}{2}} \qquad (3)$$

$$CV = \sigma_i / \overline{E_i} \qquad (4)$$

In equation (3), $\sigma_i$ represents the standard deviation. In equation (4), and CV denotes the coefficient of variation.

(4) By statistically processing the results of two rounds of expert consultation, we obtained the results of the statistical parameters of each indicator in the six aspects of the user's characteristics, the implementation conditions of the parking points, the use characteristics of the parking points, the management level of the parking points, environmental sustainability, and social sustainability, as shown in Tables 3~8. Where GCR indicates the degree of confidence of the expert group in judging the level of influence of the indicator on the location of the bike-sharing.

Table 3 The importance of indexes from user's characteristics

| Category | Alternative indexes | The first round | | | The second round | | | GCR | The full mark rate |
|---|---|---|---|---|---|---|---|---|---|
| | | $\bar{E}$ | $\sigma$ | $V$ | $\bar{E}$ | $\sigma$ | $V$ | | |
| The user's characteristics | Average arrival time of users | 3.273 | 0.663 | 0.203 | 3.600 | 0.663 | 0.184 | 4.556 | 0.320 |
| | Walking distance of users to the stopping point | 4.091 | 0.671 | 0.164 | 4.500 | 0.671 | 0.149 | 4.444 | 0.640 |
| | Time urgency | 3.364 | 0.900 | 0.268 | 4.000 | 0.632 | 0.158 | 4.333 | 0.640 |
| | The user 's search time | 3.818 | 0.600 | 0.157 | 4.200 | 0.600 | 0.143 | 4.667 | 0.640 |
| | The user 's return time | 3.455 | 0.748 | 0.217 | 3.700 | 0.781 | 0.211 | 3.889 | 0.160 |

Table 4 The importance of indexes from the implementation conditions of the parking points

| Category | Alternative indexes | The first round | | | The second round | | | GCR | The full mark rate |
|---|---|---|---|---|---|---|---|---|---|
| | | $\bar{E}$ | $\sigma$ | $V$ | $\bar{E}$ | $\sigma$ | $V$ | | |
| The implementation conditions of the parking points | The flow of people around the parking point | 3.273 | 1.114 | 0.340 | 3.700 | 1.005 | 0.272 | 4.111 | 0.320 |
| | The impact on business benefits | 3.000 | 0.900 | 0.300 | 3.400 | 0.917 | 0.270 | 3.778 | 0.800 |
| | The impact on environmental benefits | 2.182 | 0.917 | 0.420 | 2.400 | 0.917 | 0.382 | 4.000 | 0.000 |
| | The impact on the surrounding traffic environment | 2.909 | 0.980 | 0.337 | 3.200 | 0.980 | 0.306 | 4.000 | 0.160 |

Table 5 The importance of indexes from the use characteristics of the parking points

| Category | Alternative indexes | The first round | | | The second round | | | GCR | The full mark rate |
|---|---|---|---|---|---|---|---|---|---|
| | | $\bar{E}$ | $\sigma$ | $V$ | $\bar{E}$ | $\sigma$ | $V$ | | |
| The use characteristics of the parking points | The area of land occupied | 2.909 | 1.327 | 0.456 | 3.200 | 1.327 | 0.415 | 4.111 | 0.480 |
| | The turnover rate | 3.091 | 0.663 | 0.215 | 3.300 | 0.640 | 0.194 | 4.000 | 0.320 |
| | The usage rate | 3.636 | 0.632 | 0.174 | 3.900 | 0.700 | 0.179 | 4.222 | 0.640 |
| | Coverage | 2.273 | 1.025 | 0.451 | 2.900 | 0.831 | 0.286 | 4.444 | 0.160 |
| | Maximum parking capacity | 2.545 | 0.748 | 0.294 | 2.800 | 0.748 | 0.267 | 4.333 | 0.160 |
| | Maximum bicycle demand | 3.273 | 1.356 | 0.414 | 4.200 | 0.600 | 0.143 | 4.000 | 0.800 |
| | Minimum service distance | 3.455 | 0.872 | 0.252 | 3.800 | 0.748 | 0.197 | 4.333 | 0.640 |
| | Service Radius | 3.091 | 1.020 | 0.330 | 3.800 | 0.600 | 0.158 | 4.222 | 0.640 |
| | Distance to the entrance/exit | 2.182 | 0.800 | 0.367 | 2.400 | 0.800 | 0.333 | 4.444 | 0.160 |
| | Average usage | 2.545 | 0.872 | 0.342 | 2.900 | 0.831 | 0.286 | 4.111 | 0.000 |
| | Quality of the mobile network | 3.545 | 1.136 | 0.320 | 4.000 | 1.095 | 0.274 | 4.556 | 0.800 |

Table 6 The importance of indexes from the management level of the parking points

| Category | Alternative indexes | The first round | | | The second round | | | GCR | The full mark rate |
|---|---|---|---|---|---|---|---|---|---|
| | | $\bar{E}$ | $\sigma$ | $V$ | $\bar{E}$ | $\sigma$ | $V$ | | |
| The management level of the parking points | The strength of national policy | 3.000 | 1.269 | 0.423 | 3.300 | 1.269 | 0.385 | 4.556 | 0.320 |
| | The improvement of enterprise management measures | 3.091 | 0.917 | 0.297 | 3.300 | 0.900 | 0.273 | 4.111 | 0.320 |
| | Campus management facilities | 3.909 | 0.640 | 0.164 | 4.100 | 0.700 | 0.171 | 4.556 | 0.480 |

Table 7 The importance of indexes from environmental sustainability

| Category | Alternative indexes | The first round | | | The second round | | | GCR | The full mark rate |
|---|---|---|---|---|---|---|---|---|---|
| | | $\bar{E}$ | $\sigma$ | $V$ | $\bar{E}$ | $\sigma$ | $V$ | | |
| Environmental Sustainability | Landscape integration | 2.455 | 0.900 | 0.367 | 2.700 | 0.900 | 0.333 | 4.111 | 0.000 |
| | The compatibility between parking points and other parking space areas | 3.091 | 0.917 | 0.297 | 3.700 | 0.640 | 0.173 | 4.000 | 0.640 |
| | The degree of optimization of the parking order | 3.455 | 0.600 | 0.174 | 3.800 | 0.600 | 0.158 | 3.889 | 0.640 |

Table 8 The importance of indexes from social sustainability

| Category | Alternative indexes | The first round | | | The second round | | | GCR | The full mark rate |
|---|---|---|---|---|---|---|---|---|---|
| | | $\bar{E}$ | $\sigma$ | $V$ | $\bar{E}$ | $\sigma$ | $V$ | | |
| Social Sustainability | The construction of enterprise environmental protection culture | 3.273 | 0.800 | 0.244 | 3.700 | 0.640 | 0.173 | 4.333 | 0.640 |
| | The propaganda of campus environmental protection concept | 2.909 | 0.980 | 0.337 | 3.200 | 0.980 | 0.306 | 4.000 | 0.320 |
| | The implementation efforts of environmental education on campus | 3.091 | 0.800 | 0.259 | 3.700 | 0.640 | 0.173 | 4.222 | 0.640 |
| | The environmental protection concept of users | 3.455 | 0.872 | 0.252 | 4.100 | 0.539 | 0.131 | 3.778 | 0.640 |

Based on the statistical results, the standard deviation and coefficient of variation of the indicators in the statistical results derived from the 2nd round of consultation are smaller than those in the 1st round. This phenomenon indicated that the dispersion of expert judgment results in the 2nd round of consultation is smaller, and the degree of coordination is higher. And the GCRs of the experts' confidence in the judgment of the evaluation indexes are all greater than

3, indicating that the consultation results are closer to the actual situation of the experts' subjective assessment. SPSS analysis of the counseling results yielded a significance test of $P<0.01$, indicating that the results of this round of counseling were desirable.

*2.2.3 Determination of the evaluation system*

After rigorous statistical analysis and based on the selection criteria, the following secondary evaluation indicators were determined.

(1) Three secondary evaluation indicators related to the users were determined, including walking distance of users to the stopping point, time urgency, and the user's search time. Firstly, the primary target users of campus bike-sharing are students. The proximity of bike-sharing station locations to key points such as academic buildings, student dormitories, and dining areas significantly influences students' utilization of bike-sharing services. Moreover, time urgency is a significant factor influencing the utilization of bike-sharing services. Surveys indicate that lateness to classes is a prevalent issue among college students, and students demonstrate distinct patterns of time urgency in their travel behavior. The level of time urgency directly affects their choice of transportation, leading to the selection of various modes of transportation(Jiageng et al., 2022). Lastly, the user's search time is also a crucial factor in evaluating the quality of bike-sharing station locations. Close proximity and easy accessibility of bike-sharing location are essential guarantees for increasing the utilization rate of bike-sharing services.

(2) One secondary evaluation indicator related to the implementation conditions of the bike-sharing location has been determined, namely the impact of business benefits. When companies deploy bike-sharing services on campus, one of their objectives is to generate profits while facilitating convenient transportation for faculty and students. From the perspective of the company, the impact of bike-sharing location selection on enterprise benefits is an important indicator for evaluating the implementation conditions of the stations. Companies should focus on enhancing and improving stations that are financially beneficial. Simultaneously, while considering the convenience of faculty and students, underperforming stations should be either discontinued or optimized. This approach will help to increase the utilization rate of bike-sharing services.

(3) Five secondary evaluation indicators related to the usage characteristics of bike-sharing stations have been determined, including the usage rate, maximum bicycle demand, minimum service distance, service radius, and the quality of mobile network. The usage rate and maximum bicycle demand are the most direct indicators that reflect the rationality of bike-sharing station establishment. They serve as crucial indicators for evaluating the scalability of location selection and the satisfaction of users' utilization. The minimum service distance refers to the minimum reachable distance for users when using bike-sharing services. A larger minimum service distance corresponds to higher user satisfaction and better service quality. The service radius refers to the spatial or temporal distance within which bike-sharing services can be accessed and utilized. The quality of mobile network connectivity and the accuracy of the app's positioning directly affect the satisfaction level of bike-sharing users when using the mobile app for scanning QR codes to unlock bikes and for returning them.

Due to the relatively fewer secondary indicators related to the implementation conditions of bike-sharing stations, this study combines the implementation conditions and use characteristics of the stations into one criterion level when establishing the evaluation system for bike-sharing site selection.

(4) Two secondary evaluation indicators related to environmental sustainability have been determined, including the compatibility between parking points and other parking space areas and the degree of optimization of the parking order. Firstly, bike-sharing stations in universities are primarily located in areas such as roadside sidewalks and vacant spaces near buildings. As students and faculty have various transportation options to choose from, the compatibility between shared bicycle docking points and parking spaces for other modes of transportation directly impacts the satisfaction and convenience of parking and retrieving bicycles. Consequently, it impacts the environmental and operational benefits of land utilization. In the case of roadside sidewalks and vacant spaces near buildings, although they offer convenient access for parking facilities, the parking is often irregular and occupies a large area. This not only hampers the maintenance of the surrounding landscape but also has the most significant impact on the campus environment. Therefore, he degree of optimization of the parking order at bike-sharing stations is an important indicator that affects the sustainability of the campus environment.

(5) Three secondary evaluation indicators related to social sustainability were determined, including, the construction of campus environmental protection culture, the implementation efforts of environmental education on campus, and the environmental protection concept of users. Firstly, the construction of campus environmental protection culture refers to the promotion of environmental awareness by campus through slogans on the promotional board. This aims to deepen the understanding of environmental protection among faculty and students and enhance the social sustainability of campus transportation. Secondly, the implementation efforts of environmental education on campus refers to the frequency and intensity of environmental protection promotion and education carried out by the school's library, such as environmental books, bulletin boards, radio stations, and other resources related to sustainable development. It also includes the presence of environmental slogans and the level of development of environmental facilities within the campus. Lastly, the environmental protection concept of users plays a crucial role in promoting sustainable development in campus transportation through the use of bike-sharing services. The adoption of environmentally friendly attitudes and behaviors by students when using shared bicycles is of significant importance in fostering sustainable practices and creating a positive impact on the overall transportation system of the school.

The evaluation indexes of campus bike-sharing location selection were finally determined through a comprehensive analysis of the modified Delphi method of screening indexes. In this paper, we classify and analyze the location of shared bicycle parking points in 4 aspects: the user's characteristics, the implementation and use characteristics of the parking points, environmental sustainability, and social sustainability, and propose 14 evaluation indicators in the indicator layer based on the guideline layer from the interests of 4 parties: users, implementers, planning managers, and the environment, respectively, as shown in Fig.3 below.

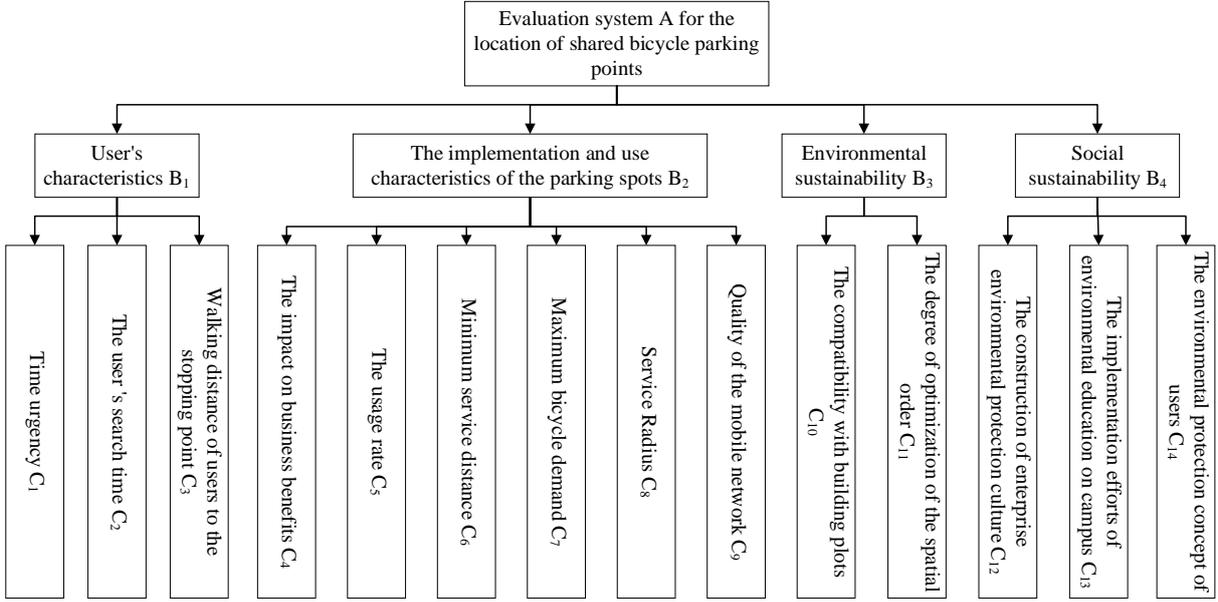

Fig.3 Evaluation system for the location of bike-sharing parking points

## 2.3 Evaluation Methods

In order to accurately evaluate the relative importance and priority of each indicator, it is crucial to assign weights within the evaluation system. By assigning weights to the evaluation indicators, we can explicitly reflect the significance of each indicator in the overall assessment process. This helps in avoiding a one-size-fits-all approach and enables a more tailored and accurate evaluation. Therefore, this chapter first introduces the methods for determining weights, including the determination of subjective weights, objective weights, and comprehensive weights. It then discusses the application of the fuzzy comprehensive evaluation method.

*2.3.1 Determination of weight*

In the determination of weights, both subjective and objective approaches have their own set of advantages and disadvantages. One of the advantages of subjective weights is their ability to incorporate the preferences and insights of relevant stakeholders, allowing for a comprehensive and inclusive decision-making process. Subjective weights have the capacity to take into account qualitative factors, expert knowledge, and nuanced contextual nuances that might be difficult to quantify. Nevertheless, it is important to note that subjective weights are vulnerable to biases and individual perceptions, which can potentially result in inconsistent or unfair outcomes. On the contrary, objective weights offer a systematic and transparent approach that relies on quantitative data and mathematical models, minimizing personal biases and promoting a more objective evaluation. However, it is important to acknowledge that objective weights may overlook qualitative aspects and heavily depend on available data and the accuracy of the chosen model. Taking into account these strengths and limitations, this research embraces a comprehensive approach that integrates both subjective and objective weights. The purpose of this combination is to capitalize on the strengths of each method while mitigating their individual weaknesses. Through the integration of subjective insights with the systematic foundation of objective weights, a more robust and well-informed evaluation system can be established, resulting in improved accuracy, transparency, and comprehensiveness in determining comprehensive weights.

*2.3.1.1 Determination of subjective weight*

In this study, the Analytic Hierarchy Process (AHP) is used to calculate the subjective weights of the indicators. AHP is a widely employed decision-making analysis methodology utilized to tackle complex problems. By integrating qualitative and quantitative elements, it facilitates the breakdown of intricate problems into their fundamental components and establishes a hierarchical structure based on their relative importance. Through pairwise comparisons, it determines the weights assigned to each criterion, thereby providing valuable assistance in the decision-making process.

(1) Construction of Judgment Matrix

In order to establish the weights for each criterion, it is essential to conduct pairwise comparisons between all the criteria. This process of pairwise comparison alleviates the challenge of directly comparing different criteria and enhances the precision of the weight assignments. Within the judgment matrix, the relative superiority or inferiority of criteria is assessed using a scale of importance. A higher numerical value signifies a greater level of importance, whereas a lower value indicates a lower level of importance. The assessment and rating of the criteria are conducted by bike-sharing planning managers, university teachers specializing in transportation planning and management, and relevant professional students. The scoring criteria for the 1-9 scale are provided in Table 9. Following this, the judgment matrix X is constructed.

$$X = \left(a_{ij}\right)_{m \times n} = \begin{bmatrix} a_{11} & \cdots & a_{1n} \\ \vdots & \ddots & \vdots \\ a_{m1} & \cdots & a_{mn} \end{bmatrix} \qquad (5)$$

Table 9 Scale of Importance and its Meaning

| Scale value | Definition (Comparison between two criteria) |
| --- | --- |
| 1 | Equally important |
| 3 | The former is slightly more important than the latter. |
| 5 | The former is significantly more important than the latter. |
| 7 | The former is strongly more important than the latter. |
| 9 | The former is extremely more important than the latter. |
| 2,4,6,8 | Intermediate value of the above judgment |

(2) Consistency check

Determine the maximum eigenvalue $\lambda_{max}$ of the judgment matrix and its corresponding eigenvector $\alpha$. Calculate the eigenvalue $\lambda_{max}$:

$$\lambda_{max} = \frac{1}{n} \sum_{i=1}^{n} \frac{(A\alpha)_i}{n\alpha_i} \qquad (6)$$

Where n represents the order (or dimension) of the judgment matrix.

The Consistency Index (CI) is defined as follows:

$$CI = \frac{\lambda_{max} - n}{n - 1} \qquad (7)$$

The average random index (RI) is provided in Table 10:

Table 10 The Average Random Index (RI)

| n | 1 | 2 | 3 | 4 | 5 | 6 | 7 | 8 | 9 |
|---|---|---|---|---|---|---|---|---|---|
| RI | 0 | 0 | 0.58 | 0.90 | 1.12 | 1.24 | 1.32 | 1.41 | 1.45 |

When the Consistency Ratio (CR) is less than 0.10, it indicates that the constructed judgment matrix is consistent. Otherwise, the judgment matrix should be appropriately revised until the Consistency Ratio meets the required condition.

$$CR = \frac{CI}{RI} \qquad (8)$$

*2.3.1.2 Determination of objective weight*

In this study, the entropy weight method is employed to determine the objective weights of the indicators. As an objective weighting method, the Entropy Weight Method calculates the entropy weight for each criterion based on the variation degree among the criteria using information entropy. These entropy weights are subsequently utilized to adjust the criterion weights, resulting in a more objective assessment of the criterion weights during practical implementation. In this study, the Entropy Weight Method is utilized to objectively determine the weights of campus bike-sharing location based on relevant data.

(1) Obtaining the qualitative criterion evaluation matrix by collecting scores from relevant personnel in the bike-sharing field.

We invite planning and management experts to evaluate the secondary indicators within the evaluation system. The six sub-indicators under the primary criterion 'The implementation and use characteristics of the location' are evaluated based on three evaluation criteria: 'Excellent Location', 'Good Location', and 'Poor Location'. In this study, the qualitative evaluation scores for the qualitative evaluation indicators are derived by consolidating the percentage values chosen according to a specific evaluation criterion. These scores are subsequently utilized to generate a comprehensive set of evaluation comments.

(2) By employing membership functions, a set of evaluation comments is generated for the quantitative evaluation indicators.

(3) Standardize the original data matrix $X_{ij}$ and compute the relative importance weight $p_{ij}$ of the jth sub-indicator for the ith shared bicycle docking station:

$$p_{ij} = \frac{X_{ij}}{\sum_{i=1}^{m} X_{ij}} \tag{9}$$

(4) Calculate the information entropy $e_j$ for the ith criterion.

$$e_j = -\frac{1}{\ln n} \sum_{i=1}^{n} p_{ij} \ln p_{ij} \tag{10}$$

(5) Calculate the entropy weight $w_j$ for the ith criterion.

$$w_j = \frac{1-e_j}{\sum_{j=1}^{m}(1-e_j)}, j=1,2,\cdots,14 \tag{11}$$

*2.3.1.3 Integration of comprehensive weight*

The AHP takes into consideration the knowledge and experience of experts, as well as the intentions and preferences of decision-makers. Although the prioritization of criterion weights often exhibits a high level of rationality, it still cannot completely overcome the significant subjective bias inherent in the process. The entropy weight method effectively explores the information inherent in the original data, resulting in relatively objective outcomes. However, it does not reflect the knowledge and experience of experts or the opinions of decision-makers. As a result, the obtained weights may not always align with the actual importance levels and may even contradict them in some cases. Analyzing the advantages and disadvantages of both methods, this study combines the results of the AHP and the entropy weight method to obtain a comprehensive indicator weight vector $\omega = (\omega_j)_{1 \times n}$ that takes into account both subjective and objective factors.

$$\omega_j = \alpha \omega_j^{'} + (1-\alpha)\omega_j^{"}, 0 \le \alpha \le 1 \tag{12}$$

The comprehensive weight varies with the change of parameter $\alpha$, where $\alpha$ takes the values of 1 and 0, corresponding to the AHP method and the entropy weight method, respectively. After conducting a literature review(Mon et al., 1994) and considering the practical aspects of evaluating the location of shared bicycle docking stations in this study, it is determined that the optimal value for parameter $\alpha$ is 0.5. This value leads to the most suitable comprehensive weight, as it strikes a balanced consideration between the AHP method and the entropy weight method.

$$\omega_j = 0.5\omega_j^{'} + 0.5\omega_j^{"} \tag{13}$$

*2.3.2 Fuzzy comprehensive evaluation method*

Due to the ability of the fuzzy comprehensive evaluation method to effectively address the subjective initiative and objective ambiguity in the evaluation process, this study utilizes the fuzzy comprehensive evaluation method to analyze the location schemes of bike-sharing docking stations on campus. This analysis is conducted based on the utilization of the AHP and entropy weight method to obtain both subjective and objective weights.

$$B_i = W_i \bullet R_i = \begin{bmatrix} B_1 \\ B_2 \\ \vdots \\ B_i \end{bmatrix} \quad (14)$$

Where $B_i$ represents the fuzzy operation result of all second-level indicators covered by the ith variable factor in layer $B$, $W_i$ represents the weights of all second-level indicators corresponding to the ith variable factor in layer $B$, and $R_i$ represents the fuzzy evaluation matrix.

We will choose the plan associated with $B_m = max\{B_1, B_2, \cdots, B_i\}$ as the optimal location for bike-sharing docking stations.

## 3. Case study

Recognizing the significance of robust sampling procedures in bolstering the credibility of research outcomes, this study adopted a well-considered sampling frame and strategy during the case verification process. The selection of Henan Polytechnic University as the case study site was based on several considerations that align with established research norms and objectives. Based on the fact that Henan Polytechnic University is among the top 100 universities in China in terms of land area and exhibits typical characteristics of a general university, this study focuses on the bike-sharing location selection in the South Campus of Henan Polytechnic University. The constructed evaluation system for campus bike-sharing site selection is validated and analyzed through a case study in this specific location. The choice of Henan Polytechnic University was made due to its representative nature as a typical higher education institution with diverse stakeholder groups and mobility patterns. Furthermore, the university's location within a dynamic urban setting presented a microcosm of the urban challenges and opportunities that the proposed site evaluation model aimed to address. This deliberate selection aimed to enhance the external validity of the study, enabling extrapolation of findings to broader urban contexts.

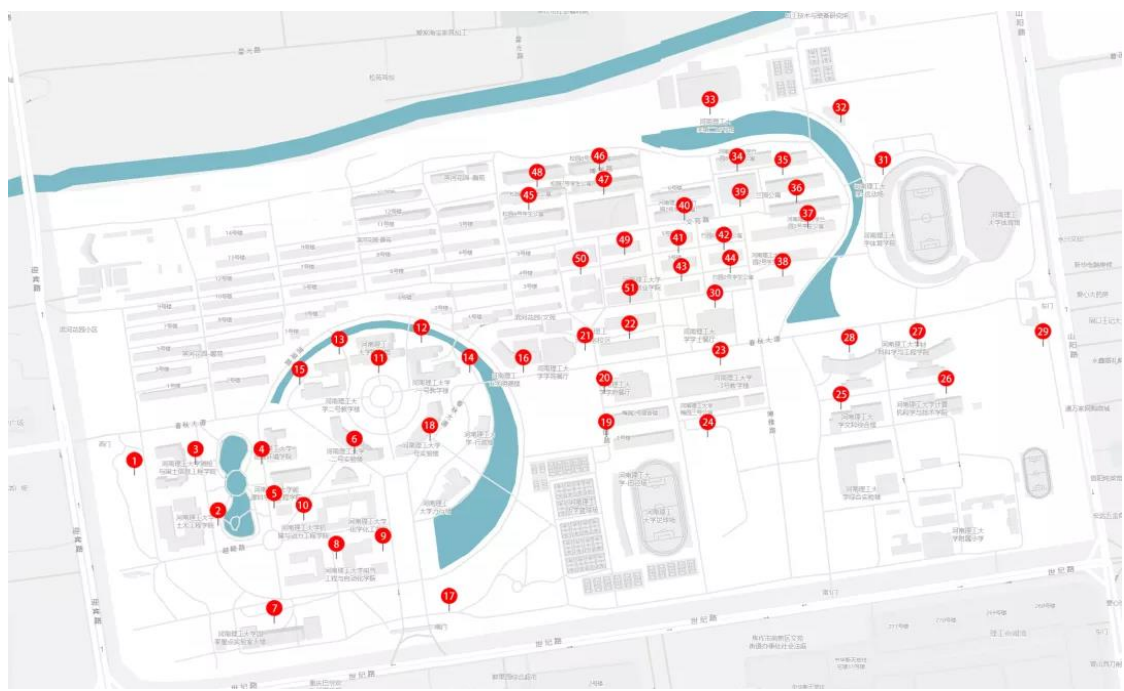

Fig.4 Distribution of shared bicycle stations

As shown in Fig.4, this paper first conducted a field survey of 45 stops on this campus, which found that most of the stops were concentrated around teaching, recreational, and other facilities with the high pedestrian flow for the convenience of students using bicycles. The size and number of bicycles parked at each stop on campus vary widely, with a few dozen shared bicycles arranged at a few docks but only a few bicycles at others. The problem is that some of the stops on campus are far from the target building and unreasonably set. There are no parking points around many restaurants, and students who ride shared bikes to restaurants need to be fully considered. The survey found that there are more problems in the shared bicycle parking points on the south campus of Henan Polytechnic University. So firstly, this paper uses the analytic hierarchy process and entropy weight method to evaluate each parking point comprehensively and makes corresponding improvements according to the evaluation results.

### *3.1 Determination of comprehensive weight*

### *3.1.1 Determination of subjective weight based on AHP*

Firstly, we construct the judgment matrix of the total criterion layer for evaluating the location of the target shared bicycle stop A, as shown in Table 11.

Table 11 Target layer A judgment matrix

| A | $B_1$ | $B_2$ | $B_3$ | $B_4$ |
|---|---|---|---|---|
| $B_1$ | 1 | 3 | 3 | 3 |
| $B_2$ | 1/3 | 1 | 3 | 3 |
| $B_3$ | 1/3 | 1/3 | 1 | 1 |
| $B_4$ | 1/3 | 1/3 | 1 | 1 |

After the consistency test of the judgment matrix, it is calculated that CR=0.0592<0.1, which indicates that the consistency test passes.

Secondly, the evaluation factor judgment matrices for the general criterion layers B1, B2, B3 and B4 are constructed respectively, as shown in Tables 12~15.

Table12 $B_1$ judgment matrix

| $B_1$ | $C_1$ | $C_2$ | $C_3$ |
|---|---|---|---|
| $C_1$ | 1 | 3 | 3 |
| $C_2$ | 1/3 | 1 | 1/2 |
| $C_3$ | 1/3 | 2 | 1 |

Table13 $B_2$ judgment matrix

| $B_2$ | $C_4$ | $C_5$ | $C_6$ | $C_7$ | $C_8$ | $C_9$ |
|---|---|---|---|---|---|---|
| $C_4$ | 1 | 1/2 | 1/3 | 1/3 | 1/3 | 1/3 |
| $C_5$ | 2 | 1 | 1/2 | 1/2 | 1/2 | 1/2 |
| $C_6$ | 3 | 2 | 1 | 2 | 1 | 1 |
| $C_7$ | 3 | 2 | 1/2 | 1 | 1/2 | 1/2 |
| $C_8$ | 3 | 2 | 1 | 2 | 1 | 1 |
| $C_9$ | 3 | 2 | 1 | 2 | 1 | 1 |

Table14 $B_3$ judgment matrix

| $B_3$ | $C_{10}$ | $C_{11}$ |
|---|---|---|
| $C_{10}$ | 1 | 1/3 |
| $C_{11}$ | 3 | 1 |

Table15 $B_4$ judgment matrix

| $B_4$ | $C_{12}$ | $C_{13}$ | $C_{14}$ |
|---|---|---|---|
| $C_{12}$ | 1 | 1/3 | 1/3 |
| $C_{13}$ | 3 | 1 | 1/2 |
| $C_{14}$ | 3 | 2 | 1 |

After the consistency test of the above judgment matrix, it is calculated that CR = 0.0596,0.0147,0,0.0596 are all less than 0.1, which indicates that the consistency test is passed and the results are reasonable and effective.

After the column vector normalization operation of the evaluation factors in the judgment matrix, the weight value of each evaluation factor for target A is calculated, as shown in Table 16.

Table 16 Subjective weight of evaluation factor for shared bicycle parking points

| Goal layer A | Criterion layer B | Weight | Evaluation factor layer C | Relative weight | Weight |
|---|---|---|---|---|---|
| Location evaluation of shared bicycle parking points | $B_1$ | 0.487 | $C_1$ | 0.594 | 0.289 |
| | | | $C_2$ | 0.157 | 0.076 |
| | | | $C_3$ | 0.249 | 0.121 |
| | $B_2$ | 0.276 | $C_4$ | 0.065 | 0.018 |
| | | | $C_5$ | 0.107 | 0.030 |
| | | | $C_6$ | 0.227 | 0.063 |
| | | | $C_7$ | 0.147 | 0.041 |
| | | | $C_8$ | 0.227 | 0.063 |
| | | | $C_9$ | 0.227 | 0.063 |
| | $B_3$ | 0.118 | $C_{10}$ | 0.250 | 0.030 |
| | | | $C_{11}$ | 0.750 | 0.089 |
| | $B_4$ | 0.118 | $C_{12}$ | 0.140 | 0.017 |
| | | | $C_{13}$ | 0.333 | 0.040 |
| | | | $C_{14}$ | 0.528 | 0.062 |

*3.1.2 Determination of objective weight based on entropy weight method*

The membership data of all quantitative indicators are summarized into a fuzzy matrix, as shown in Table 17.

The field survey data are calculated by Equations (5)-(7), and the objective weight of the secondary index evaluation factor of the bike-sharing parking point location shown in Table 18 is obtained.

Table 17 Evaluation criteria of quantitative index

| Criterion layer B | Evaluation factor layer C | Excellent | Good | Poor |
|---|---|---|---|---|
| $B_1$ | $C_1$ | 0.1 | 0.5 | 0.4 |
| | $C_2$ | 0.3 | 0.6 | 0.1 |
| | $C_3$ | 0.2 | 0.5 | 0.3 |
| $B_2$ | $C_4$ | 0 | 0.5 | 0.5 |
| | $C_5$ | 0 | 0.4 | 0.6 |
| | $C_6$ | 0.1 | 0.3 | 0.6 |
| | $C_7$ | 0.4 | 0.4 | 0.2 |
| | $C_8$ | 0.4 | 0.4 | 0.2 |
| | $C_9$ | 0 | 0.2 | 0.8 |
| $B_3$ | $C_{10}$ | 0.2 | 0.4 | 0.4 |
| | $C_{11}$ | 0.1 | 0.6 | 0.3 |
| $B_4$ | $C_{12}$ | 0.3 | 0.5 | 0.2 |
| | $C_{13}$ | 0.4 | 0.6 | 0 |
| | $C_{14}$ | 0.3 | 0.6 | 0.1 |

Table 18 Objective weight of evaluation factor for shared bicycle parking points

| Goal layer A | Criterion layer B | Weights | Evaluation factor layer C | Objective weights |
|---|---|---|---|---|
| Location evaluation of shared bicycle parking points | $B_1$ | 0.2081 | $C_1$ | 0.0460 |
| | | | $C_2$ | 0.1313 |
| | | | $C_3$ | 0.0308 |
| | $B_2$ | 0.533 | $C_4$ | 0.3449 |
| | | | $C_5$ | 0.0190 |
| | | | $C_6$ | 0.0441 |
| | | | $C_7$ | 0.0443 |
| | | | $C_8$ | 0.0398 |
| | | | $C_9$ | 0.0409 |
| | $B_3$ | 0.0457 | $C_{10}$ | 0.0138 |
| | | | $C_{11}$ | 0.0319 |
| | $B_4$ | 0.2132 | $C_{12}$ | 0.1563 |
| | | | $C_{13}$ | 0.0360 |
| | | | $C_{14}$ | 0.0209 |

*3.1.3 Determination of comprehensive weight*

The subjective weight of the evaluation index of the shared bicycle parking site location calculated by the analytic hierarchy process in table 15 and the objective weight of the evaluation index of the shared bicycle parking site location calculated by the entropy weight method in table 19 are obtained. The comprehensive weight shown in table 18 is calculated by equation (9).

Table 19 The comprehensive weight of evaluation factor for shared bicycle parking points

| Goal layer A | Criterion layer B | Weights | Evaluation factor layer C | Subjective weight | Objective weights | Comprehensive weight |
|---|---|---|---|---|---|---|
| Location evaluation of shared bicycle parking points | $B_1$ | 0.348 | $C_1$ | 0.289 | 0.0460 | 0.1675 |
| | | | $C_2$ | 0.076 | 0.1313 | 0.10365 |
| | | | $C_3$ | 0.121 | 0.0308 | 0.0759 |
| | $B_2$ | 0.405 | $C_4$ | 0.018 | 0.3449 | 0.18145 |
| | | | $C_5$ | 0.030 | 0.0190 | 0.0245 |
| | | | $C_6$ | 0.063 | 0.0441 | 0.05355 |
| | | | $C_7$ | 0.041 | 0.0443 | 0.04265 |
| | | | $C_8$ | 0.063 | 0.0398 | 0.0514 |
| | | | $C_9$ | 0.063 | 0.0409 | 0.05195 |
| | $B_3$ | 0.082 | $C_{10}$ | 0.030 | 0.0138 | 0.0219 |
| | | | $C_{11}$ | 0.089 | 0.0319 | 0.06045 |
| | $B_4$ | 0.166 | $C_{12}$ | 0.017 | 0.1563 | 0.08665 |
| | | | $C_{13}$ | 0.040 | 0.0360 | 0.038 |
| | | | $C_{14}$ | 0.062 | 0.0209 | 0.04145 |

## *3.2 Fuzzy comprehensive evaluation*

### *3.2.1 First-level evaluation index fuzzy evaluation*

According to Equation (10), the comprehensive evaluation results in Table 20 were obtained by the operation of the matrix.

Table 20 First-level evaluation results

| Criterion layer B | Excellent | Good | Poor |
|---|---|---|---|
| User's characteristics $B_1$ | 0.1563 | 0.5157 | 0.328 |
| The implementation and use characteristics of the parking points $B_2$ | 0.1723 | 0.3384 | 0.4893 |
| Environmental sustainability $B_3$ | 0.125 | 0.55 | 0.325 |
| Social sustainability $B_4$ | 0.3326 | 0.5866 | 0.0808 |

### *3.2.2 Secondary evaluation index fuzzy evaluation*

According to the results of the first-level evaluation, the results of the second-level evaluation of the location of the campus shared bicycle parking point are obtained through the formula (10). B refers to the level of site selection evaluation obtained for the goal layer A, and W refers to the weight relative to the evaluation target of $B_1$, $B_2$, $B_3$ and $B_4$. From the above, it can be seen that:

$$W = [0.348, 0.405, 0.082, 0.166]$$

R is the evaluation matrix. From the first-level evaluation results above, it can be seen that:

$$R = \begin{bmatrix} 0.1563 & 0.5157 & 0.328 \\ 0.1723 & 0.3384 & 0.4893 \\ 0.125 & 0.55 & 0.325 \\ 0.3326 & 0.5866 & 0.0808 \end{bmatrix}$$

We can obtain the results of the secondary comprehensive evaluation using the formula (10), as shown in Table 21.

Table 21 Secondary evaluation results

| Campus shared bicycle parking location | Evaluation result | | |
|---|---|---|---|
| | Excellent | Good | Poor |
| | 0.1896 | 0.458 | 0.3524 |

The evaluation results of the location of shared bicycle parking points on campus show that the possibilities of ' excellent, '' good, 'and ' poor ' are 18.96%、45.8%, and 35.24%, respectively. Based on the principle of maximum membership degree, the evaluation result of the location of shared bicycle parking points at Henan Polytechnic University is 45.8 %. Therefore, the evaluation result of the location of shared bicycle parking points in Henan Polytechnic University is ' good. '.

## 4. Conclusion

In the pursuit of advancing sustainable transportation solutions, this study aimed to create a comprehensive evaluation framework rooted in the principles of sustainable development for selecting optimal locations for campus bike-sharing programs. The overarching goal was to offer decision-makers a systematic approach that holistically considers a range of factors, ultimately enhancing the effectiveness and long-term sustainability of bike-sharing initiatives. Through the integration of various methodologies and empirical analysis, this research has yielded valuable insights that extend beyond immediate implementation considerations, offering broader implications for the realm of sustainable urban mobility and policy-making.

Our investigation harnessed an improved Delphi method to identify and distill 14 influential factors influencing location selection for campus bike-sharing. These factors spanned domains of operational efficiency, environmental impact, and user accessibility, and social impact. The subsequent application of the analytic hierarchy process (AHP) and the entropy weight method facilitated the assignment of relative weights to these factors, acknowledging their varied significance within the decision-making context. Employing the fuzzy comprehensive evaluation method further enabled the quantitative assessment of location quality. The outcomes of our analysis highlighted several paramount factors, including time urgency, the impact of business benefits, the degree of optimization of the spatial order, and the construction of enterprise environmental protection culture. These findings go beyond the immediate scope of location selection, offering a nuanced understanding of the multifaceted interplay between operational, economic, and environmental dimensions within bike-sharing initiatives. Decision-makers can leverage this nuanced insight to optimize program design, resource allocation, and stakeholder engagement, consequently fostering more sustainable and resilient campus transportation systems. The practical validation of our model through the case study conducted at Henan Polytechnic University underscores the real-world applicability of the evaluation framework. The alignment between the proposed model and the existing bike-sharing infrastructure on the South Campus affirms its utility and adaptability. However, while the current study delves into campus-specific considerations, the insights gained can be extrapolated to inform broader urban planning efforts. By analyzing campus settings as microcosms of urban complexities, this research contributes to the broader discourse on sustainable urban mobility, offering transferable insights for municipalities and policymakers grappling with similar transportation challenges.

Looking ahead, the study's exploration of a comprehensive evaluation method for campus bike-sharing locations lays the groundwork for future research endeavors. One potential avenue is the expansion of the site selection problem from single-land nature contexts to encompass complex land use scenarios. This broader approach can provide a more encompassing evaluation index system that reflects the intricate interactions between shared bicycle systems, urban land use, and social dynamics. Such an endeavor would not only deepen our understanding of sustainable mobility but also offer tailored solutions that balance the interests of stakeholders, the environment, and shared bicycle enterprises. In conclusion, this study's development of a location selection evaluation system for campus bike-sharing programs contributes not only to informed decision-making within campus settings but also to the broader advancement of sustainable transportation alternatives. This work stands as a foundational step toward cultivating more environmentally conscious and efficient transportation systems, resonating far beyond campus boundaries.

**Acknowledgements**

We want to express our gratitude to anonymous referees for their valuable and constructive comments. This research was supported by grants from Henan Science and Technology Public